# Applications and Societal Implications of Artificial Intelligence in Manufacturing: A Systematic Review

John P. Nelson
*School of Public Policy*
*Georgia Institute of Technology*
Atlanta, USA
ORCID: 0000-0002-3010-2046
jnelson330@gatech.edu

Justin B. Biddle
*Ethics, Technology, and Human Interaction Center (ETHIC$^x$)*
*School of Public Policy*
*Georgia Institute of Technology*
Atlanta, USA
ORCID: 0000-0002-8149-5759
justin.biddle@pubpolicy.gatech.edu

Philip Shapira
*Manchester Institute of Innovation Research, University of Manchester*
Manchester, UK,
and
*School of Public Policy*
*Georgia Institute of Technology*
Atlanta, USA
ORCID: 0000-0003-2488-5985
pshapira@gatech.edu

*Abstract*—This paper undertakes a systematic review of relevant extant literature to consider the potential societal implications of the growth of AI in manufacturing. We analyze the extensive range of AI applications in this domain, such as interfirm logistics coordination, firm procurement management, predictive maintenance, and shop-floor monitoring and control of processes, machinery, and workers. Additionally, we explore the uncertain societal implications of industrial AI, including its impact on the workforce, job upskilling and deskilling, cybersecurity vulnerability, and environmental consequences. After building a typology of AI applications in manufacturing, we highlight the diverse possibilities for AI's implementation at different scales and application types. We discuss the importance of considering AI's implications both for individual firms and for society at large, encompassing economic prosperity, equity, environmental health, and community safety and security. The study finds that there is a predominantly optimistic outlook in prior literature regarding AI's impact on firms, but that there is substantial debate and contention about adverse effects and the nature of AI's societal implications. The paper draws analogies to historical cases and other examples to provide a contextual perspective on potential societal effects of industrial AI. Ultimately, beneficial integration of AI in manufacturing will depend on the choices and priorities of various stakeholders, including firms and their managers and owners, technology developers, civil society organizations, and governments. A broad and balanced awareness of opportunities and risks among stakeholders is vital not only for successful and safe technical implementation but also to construct a socially beneficial and sustainable future for manufacturing in the age of AI.

## I. Introduction

Artificial intelligence (AI) encompasses a variety of flexible techniques, each with a wide variety of potential applications and effects in society. Policymakers, industry leaders, and some technicians have heralded AI, along with other "Industry 4.0" technologies such as the "Internet of Things" and cloud computing, as enabler of radical transformation within the manufacturing economy, and across society more broadly (Avis 2018; National Institute of Standards and Technology 2022; McKinsey 2022). Many firms have already adopted machine learning tools in some capacities, e.g., for product quality monitoring and factory automation, but firm leaders and analysts expect that machine learning and AI technologies will see many more (and more intensive) applications in industry as they mature (Deloitte China 2020).

AI's potential manufacturing applications range from interfirm logistics coordination, to firm procurement management, to predictive maintenance, to shop-floor monitoring and control of process flows, machines, and workers. The societal implications of industrial AI are highly uncertain; scholars and analysts have suggested both workforce gains and losses, job upskilling and deskilling, shifts in cybersecurity vulnerability, and environmental harms and benefits, among other possibilities. While there is much excitement about manufacturing AI in government and manufacturing industry literature (e.g., National Institute of Standards and Technology 2020, World Manufacturing Foundation 2020), detailed examinations of the potential societal implications of manufacturing AI are rare. This constitutes a serious gap, because construction of a prosperous and broadly beneficial future for manufacturing will require a balanced awareness among managers, policymakers, workers, and publics of the capabilities and the potential dangers and benefits of AI in manufacturing.

To inform management, governance, and further research, we review prior literature to catalog potential applications and implications of artificial intelligence in manufacturing (Appendix 1). We furthermore identify and discuss possible applications and implications omitted or not widely discussed in present AI manufacturing literature. We cite sources illustratively throughout, with comprehensive lists of sources discussing AI manufacturing applications in Appendix 2 and AI societal implications in Appendix 3.

Section 2 presents a typology of potential artificial intelligence applications in manufacturing, developed based upon prior typologies and organized by categories of *scale* and *application type*. Application scales include product, service, or tool, at the smallest scale, up to single-site operations, at the meso-scale, to intersite operations management, at the large scale. Application types include system design and development, system planning and scheduling, system tracking and monitoring, and system control. Section 3 reviews possible implications for industrial artificial intelligence for the values of individual firms and of society at large, including economic prosperity and equity, environmental health, and community safety and security. Prior literature presents a largely sunny view of AI's implications for firms, but AI's societal implications are disputed. This dispute illustrates the fact that many of AI's societal effects have yet to be determined. Section 4 reflects upon the techno-optimistic tenor of most of the reviewed literature and discusses other possible societal implications for industrial AI drawn from analogies to historical cases and from discussion of AI in other contexts. Section 5 observes that AI's effects will depend to a substantial extent upon the priorities and choices of technology developers, firms, civil society organizations, and governments, among other actors. Section 6 concludes by articulating broad programs of action by which governments, firms, and civil society may be able to shape the development and implementation of AI in industry to protect and advance societal values in danger of neglect.

## II. Artificial intelligence applications in manufacturing

### A. Typology overview

At the smallest scale, AI can be applied to the development, control, and management of products or of individual machines on the factory floor (e.g., machine tools, cutters, manufacturing robots). At the meso-scale, AI can be applied to the operations of a single manufacturing facility. At the large scale, AI can be applied to the management of multi-site supply networks, either within or across firms (Table 1). In principle, AI might also be applied to (arguably) larger-scale tasks of governance or high-order societal decision-making, but this possibility is not addressed in the reviewed literature. It should be noted, however, that whether or not AI is eventually deployed at high levels of governance, societal governance choices and political economy will affect what lower-order AI applications are developed and implemented, how, where, and with what effects.

We divide application types by the nature of the task. At all scales, AI may, first, be applied to design and development of products and systems, e.g., in predicting and assessing the performance characteristics of designs, in searching through previous designs or suggesting design characteristics, or generating designs itself. Second, it may be applied to planning and scheduling within a system design or product contract, e.g., by projecting use and inventory of materials or supplies, market supply and demand for inputs or outputs, wear on machines, or process faults requiring compensation; in assessing the possible performance of different management plans under different scenarios; or in adjusting scheduling, ordering, and process parameters in real time in response to the present states of markets, plant, or production processes. Third, AI can be applied to the tracking and monitoring of systems or products, e.g., in tracking deliveries or materials through supply chains or through production processes, machine status and performance, quality of output, worker behavior, or product status and use. Fourth, AI can be applied to process or product control, e.g., in making decisions about how to respond to process divergences, dispatching job orders or schedules, contextually communicating with workers, or in controlling manufacturing robots.

Many application types could be deployed in ensemble. For example, AI monitoring of machine status could be used to, in real time, update maintenance plans based on machine wear forecasting, and to trigger automated maintenance. Similarly, real-time monitoring of manufacturing processes and output quality could permit automated diagnosis of process faults and compensatory adjustment of process parameters. For a final example, AI could be used to plan the motions of manufacturing robots, to control robots in accordance with those plans, and to flexibly adjust motions

*Table 1. Typology of AI applications in manufacturing*

| | | Application variety | | | |
|---|---|---|---|---|---|
| | | *Design & development* | *Planning & scheduling* | *Tracking & monitoring* | *Dispatching & control* |
| **Application scale** | *Supply network* | [None in reviewed literature] | Inter-site operations coordination<br>Market forecasting | Inventory, request, & delivery tracking | Ordering & distribution |
| | *Facility* | Process parameter prediction & selection | Job scheduling & coordination<br>Maintenance planning & coordination | Energy & resource use tracking<br>Output & quality monitoring & attribution | Process reconfiguration<br>Real-time process adjustment<br>Worker instruction |
| | *Production machines & tools; Products & services* | Design property prediction & assessment<br>Generative design | Predictive error compensation<br>Forecasting usage<br>Predictive maintenance | Status monitoring<br>Use tracking<br>Feedback processing & analysis | Machine control<br>Upkeep & handling automation |

*Source:* Authors' elaboration. See Appendix 2 for comprehensive list of applications discussed in literature.
*Note:* Application scale decreases with downward movement along the y-axis.

in real time to avoid collisions with human workers or other process divergences.

It is difficult to determine the extent to which each of the applications discussed in the literature is actually occurring in industry at present, as most reviewed papers discuss both current and potential applications. However, all applications discussed here, with the possible exception of industry-wide data sharing, are at least under development. We discuss applications from smallest in scale to largest, as applications often have the most readily available concrete illustrations at small scale and become more abstract at larger ones.

*B. AI applications at the product and service level*

*1) Product and service design and development.* The use of AI in product or service development or design is widely discussed in the literature. AI could be used to predict properties of computer-modeled product designs, such as production costs and needs or structural qualities. Such prediction could facilitate design assessment and refinement (Brunton et al. 2021; Chinchanikar & Shaikh 2022; Goh et al. 2021). In a related vein, Brunton and colleagues (2021) suggest that AI could be used both to improve mathematical models of aircraft performance and to optimize the planning of aircraft trial flights, thereby reducing the total amount of real-world testing needed to evaluate new designs.

AI could also be used to suggest designs, either in whole or in part, through prediction from similar designs (Chinchanikar & Shaikh 2022; Rauch et al. 2021; Wang et al. 2020). Such designs could optimize certain process or performance characteristics (e.g., minimizing material used to achieve a certain level of component strength). AI could also offer designers suggestions for further design steps, possibly helping designers to stay abreast of design developments and even deskilling parts of the design process (Rauch et al. 2021). Such design suggestion could be extended into generative design, i.e., use of AI to generate a large number of designs for a given component or product, later to be assessed and filtered manually or through AI evaluation (Chinchanikar & Shaikh 2022; Sahoo & Lo 2022; World Manufacturing Foundation 2020). Some authors suggest that AI platforms could permit customers lacking design expertise to customize product designs in response to their needs (Goh et al. 2021; C. Liu et al. 2022; World Manufacturing Foundation 2020; Xames et al. 2022). Brunton and colleagues (2021) suggest that AI could be used to search through records of preexisting designs to avoid unnecessary design of new aircraft parts, while Nolan (2021; see also Chen 2021) discusses AI search through experiment records to identify an alloy with desired properties for 3D printing of airplane components.

Finally, AI features can be integrated into products themselves. Among our reviewed papers, only Deloitte China (2020) discusses this application. For example, Apple Face ID and Windows Hello use machine learning for facial identity authentication.

*2) Product and service planning and scheduling.* Only four papers discuss use of AI in planning and scheduling for products and services (Brunton et al. 2021; Burström et al. 2021; Kamble et al. 2022; Nguyen et al. 2022). All four suggest that AI could be used for maintenance planning through digital twinning and predictive maintenance. Digital twinning refers to maintaining customized computer models of each individual product, updated by data collection from the real product and drawing inferences about product status and behavior. Predictive maintenance entails use of models like digital twins, in combination with use and maintenance data, to foresee wear on particular product components given particular use schedules. Thus, predictive maintenance may enable more effective targeting of maintenance to product needs and reduce the need for unscheduled maintenance for complicated and long-serving systems such as aircraft. Bertolini and colleagues (2021; see also Jennings et al. 2016) suggest that AI could be used to predict obsolescence of particular products. Burström and colleagues (2021) also discuss several other applications of AI in planning products and services, including using AI to adjust contract pricing to customer risk or to customize contracts to particular clients (e.g., lowering maintenance frequency during low-product-use months).

*3) Product and service tracking and monitoring.* AI could also be applied to several problems in monitoring product status and use. Applications include tracking and processing customer or user feedback and service quality (Goh et al. 2021; Nguyen et al. 2022; Sahoo & Lo 2022; Tao et al. 2018) and tracking product use and health for maintenance (Burström et al. 2021; Kamble et al. 2022; Nguyen et al. 2022). Tracked data could be used for communicating product status to clients (Sahoo and Lo 2022), for planning future service offerings (Burström et al. 2021), and for informing the design of future products, including, potentially, by supporting the development of improved predictive models of design performance (Bruton et al. 2021).

*4) Product and service control and dispatching.* AI could be applied for product control or dispatching, though only one paper in our review set discusses this application (Brunton and colleagues 2021). This may be due to our search procedure, which focused on AI applications in the manufacturing process itself rather than downstream in product lifecycles. These authors state that AI could be used for automation of some aircraft service tasks, e.g., positioning aircraft at gates, loading or removing cargo, waste, water, and fluids, and reporting flight issues. For another example, locking or unlocking digital devices using AI facial recognition, discussed above, could also qualify as AI product control.

*C. AI applications at the manufacturing machine and tool level*

*1) Machine and tool design and development.* Machines and tools used in manufacturing are themselves manufactured products. Thus, all the uses of AI in product and service design and development discussed above may apply to manufacturing machines and tools (Kamble et al. 2022). Per Kumar (2019), AI may be particularly useful for design of custom and limited-use items such as injection molds and dies.

*2) Machine and tool planning and scheduling.* AI could be applied in several different ways to planning and scheduling of operations with individual machines or tools. One major example is predictive maintenance, as described in further detail above (Arinez et al. 2020; Huang et al. 2021; Wang et al. 2022). AI could also be used, in additive manufacturing and possibly in other applications, to predict and compensate for errors in the manufacturing process, e.g., warping during 3D printing (Meng et al. 2020; Qin et al.

2022; Sarkon et al. 2022; Xames et al. 2022). AI could also be used to plan complicated machine movements, e.g., picking and placing by robotic manipulators or robot collision avoidance (Huang et al. 2021; Li et al. 2023; National Institute of Standards and Technology 2022).

*3) Machine and tool tracking and monitoring.* In concert with planning, AI could be used to process incoming data on machines or tools to track machine status, health and performance (Arinez et al. 2020; Bertolini et al. 2021; Huang et al. 2021; Rauch et al. 2021). AI could furthermore be used to diagnose, classify, and attribute machine failures to inform management and maintenance (Arinez et al. 2020; Bertolini et al. 2021).

*4) Machine and tool control and dispatching.* AI could be used to control machine movements, potentially enabling more contextual and flexible movement, generalized programming, and self-learning for assembly (Arinez et al. 2020; Kamble et al. 2022; Li et al. 2023; National Institute of Standards and Technology 2022). These, in turn, might permit automation of more delicate and context sensitive tasks (Li et al. 2023; Z. Liu et al. 2022), reduction of programming costs, and more extensive human-robot collaboration (Huang et al. 2021; Z. Liu et al. 2022). For example, AI collision avoidance might eliminate the need for safety cages separating human workers from robots. It may also be possible to automate some maintenance tasks using AI (Li et al. 2023).

*D. AI applications at the manufacturing facility level*

*1) Manufacturing facility design and development.* AI has many potential applications in facility layout or manufacturing process design. It could be used to predict parameters of different manufacturing process designs, including input parameters such as energy use and materials use, operating parameters such as machine downtime and lead time, and performance parameters such as process duration, capacity, and yield (Bertolini et al. 2021; Huang et al. 2021; Khosravani & Nasiri 2020). This information could be used to evaluate different potential process designs and also to plan and schedule purchasing, maintenance, and process stages. Similarly, AI could be used to plan and optimize space use and retrieval routing in warehouses (World Manufacturing Report 2022). These applications are feasible both at the level of entire facilities and of particular subprocesses or cells therein. AI could be used to search within libraries of preexisting process designs to ease process design and reduce redundant effort (National Institute of Standards and Technology 2022), e.g., to select appropriate tools for use in manufacturing processes (Kumar 2019). However, such libraries are rare at present, and industry-wide design sharing like that envisioned by the U.S. National Institute of Standards and Technology (2022) does not presently occur.

*2) Manufacturing facility planning and scheduling.* In line with the above discussion of predictive maintenance, AI could be used to make and adjust predictive maintenance plans and to optimize them for particular parameters, e.g., minimization of machine downtime (Fahle et al. 2020; Huang et al. 2021; Mumali 2022). AI could also be used for scheduling and coordination of jobs (Kamble et al. 2022; Li et al. 2023). Specifically, AI may be useful for real-time rescheduling of jobs within dynamic, inconsistent, and contingent manufacturing processes (Bertolini et al. 2021; Rauch et al. 2021). AI could, similarly, be used to dynamically allocate resources within manufacturing processes (Arinez et al. 2020; Dogan & Birant 2021; Sahoo & Lo 2022). As in process design, AI could be used to compare different possible manufacturing plans along dimensions of projected performance (Huang et al. 2021). In a similar vein, AI could be used to identify points within manufacturing processes for resource-efficient quality monitoring and assessment (Brunton et al. 2021).

*3) Manufacturing facility tracking and monitoring.* AI has substantial potential applications in output quality monitoring and attribution and, relatedly, in process fault detection and attribution. Machine vision is already used for visual quality inspection and defect classification of products such as microchips. Such automated quality inspection, combined with collection of process data, may permit quality inspection at more points within production processes and upstream, real-time detection of quality problems. In combination with collection and analysis of data on machine status and behavior, AI quality control may support automated detection, classification, and attribution of process faults, which is often a difficult and labor-intensive process in large and complicated production processes (Bertolini et al. 2021; Deloitte China 2020; Qin et al. 2022).

AI could also be used for real-time tracking of energy and resource use for later analysis or for inventory planning, and for tracking and documentation of maintenance (Li et al. 2023; Tao et al. 2018; Woschank et al. 2020). Similarly, AI could be used to monitor the behavior of workers on factory floors. Applications of such monitoring could include identification of potential safety threats, or checking for worker adherence to process plans and site policies (Sira 2022; World Manufacturing Foundation 2020). AI might also be used to track how workers complete tasks in hopes of identifying more effective task methods (Brunton et al. 2021). Some authors even discuss the use of AI to formally encode formerly tacit worker skills and pass them on to new employees, preventing loss of important craft knowledge and skills when expert workers leave firms (National Institute of Standards and Technology 2022).

Additionally, AI has potential applications for site security. AI could be used to detect cyberattacks (Goh et al. 2021; Nguyen et al. 2022; World Manufacturing Foundation 2020; Xames et al. 2022).[1] However, AI could also be used for industrial espionage, e.g., to predictively reconstruct process or product designs or parameter based on incomplete data (Wang et al. 2020). Some experiments have successfully reconstructed additive manufacturing products using cellphone audio recordings of manufacturing (e.g., Hojjati et al. 2016, Faruque et al. 2016).

*4) Manufacturing facility control and dispatching.* Real-time monitoring of process data could permit AI-driven adjustment of manufacturing processes, e.g., automated rescheduling or even automated adjustment to compensate for identified process faults (Arinez et al. 2020; Bertolini et al. 2021; Kamble et al. 2022; S. Kumar et al. 2022). Similarly, AI could be used to flexibly dispatch and convey resources within manufacturing processes (Gentner et al.

---

[1] Interesting, no reviewed AI application typologies discuss the possibility of using AI to automate parts of the cyberattack process.

2022; Li et al. 2023; Tao et al. 2018). AI could be used to dispatch manufacturing jobs to production cells (Arinez and colleagues 2020; Rockwell Automation 2023). AI might also be used to communicate with workers, e.g., to convey instructions, to provide augmented reality guidance for difficult tasks, or for contextual provision of training materials (Fahle et al. 2020; Nolan 2021). In a different vein, some authors suggest that AI could be used to construct and control dynamically reconfigurable manufacturing systems capable of rapidly and easily switching between production lines (Li et al. 2023; National Institute of Standards and Technology 2022).

*E. AI applications at the supply network level*

*1) Supply network planning and scheduling.* No reviewed papers discuss use of AI to design or develop entire supply networks, perhaps because detailed, top-down design of supply networks is both difficult and rare. However, the literature does suggest that AI could be used by particular firms or agents for planning and scheduling their multi-site operations or their interactions with other firms in their supply networks (Kamble et al. 2022; National Institute of Standards and Technology 2022). AI could use historical data to project supply, demand, and pricing of production inputs and outputs for procurement and production planning (Bertolini et al. 2021; Nolan et al. 2021; World Manufacturing Foundation 2020). AI might be used to identify and target products to particular market segments (Burström et al. 2021). AI might also permit more detailed cooperation across firms. For example, AI could be used to search the production landscape for firms possessing desired production capabilities, or, with substantial data sharing, to search many different firms' records for previously used process plans (National Institute of Standards and Technology 2022).

*2) Supply network tracking and monitoring.* AI could be used to track inventory, orders, and deliveries across multiple sites (Burström et al. 2021, Deloitte China 2020, Kamble et al. 2022, Woschank et al. 2020). With firm agreement, AI could also be used to facilitate industry-wide aggregation and sharing of data on production designs, processes, and activities, e.g., through automation of collection, reporting, cleaning, organization, and intellectual property protection activities (National Institute of Standards and Technology 2022).

*3) Supply network control and dispatching.* AI could be used for placing and dispatching procurement and production orders. Such ordering would use the aforementioned AI-enabled resource, delivery, and inventory tracking and supply and demand forecasting capabilities to guide or automated dispatching of orders and resources across sites (Bertolini et

---

**Box 1. Examples of AI applications in aerospace engineering**

Brunton and colleagues (2021) discuss in detail multiple potential applications of AI in aerospace engineering and manufacturing. At the design stage, they suggest that AI could be applied to the many multidimensional optimization problems in aerospace design, and that AI could aggregate and identify patterns in lifelong aircraft service data to inform future design processes. Improved computer models of part behavior and aircraft performance could help to avoid downstream problems in assembly and operation. They also suggest that AI could be used to search and compare through aircraft components to identify preexisting suitable components and reduce redundant design and manufacturing of similar, but incompatible, parts.

In aerospace manufacturing and assembly, they cite the common applications of defect identification and prediction and fault identification and classification. Such methods can permit predictive compensation for flaws or defects. For example, aircraft assembly usually requires custom shimming, manufacturing, and insertion of small wedges to fill gaps between hull components. AI can be used to predict gaps in future builds and produce shims in advance. They also hope that, over time, computer vision observation of workers could provide data from which AI could identify best practices for difficult processes like forming flat parts into the curve of a fuselage. Such a formal knowledge base, they suggest, could eventually permit automation of these difficult processes.

Brunton and colleagues see applications for AI in aerospace verification and validation as well. They expect that AI could automate aggregation of data from test flights, which is presently a manual, time-consuming, and skill-intensive process. They also hope that AI could help to design test flight plans to get the most relevant performance data for the least amount of flight time. Last, they suggest that AI could permit collection and application of data from all tests for all performance characteristics, rather than, as in present practice, only assessing each performance characteristic through the test segment designed to investigate it. Substantial opportunities are also seen for AI-driven automation of aircraft positioning at port, waste removal, loading and unloading of passengers and luggage, fluid servicing, reporting of in-flight problems, and pre-flight checks. It is anticipated that predictive maintenance based on AI models of individual aircraft service and wear could eliminate unscheduled maintenance and the costly delays it brings.

There are several other imaginable applications of AI in aerospace engineering. AI might be applied to the design of aerospace manufacturing tools and machines as well as to the aircraft themselves. Scheduling of manufacturing tasks, allocation of factory space, and allocation and transport of resources within facilities could all potentially be automated or assisted through AI. Contextual, AI-enabled communication could instruct workers on novel processes, provide real-time guidance in delicate positioning tasks, or provide training. At the firm or supply network level, AI might be applied to track and coordinate orders, deliveries, and manufacturing tasks across sites; to predict materials shortfalls, surpluses, and input market conditions, and adjust purchasing plans; and perhaps to track and forecast aircraft demand to inform design, development, and manufacturing.

al. 2021; Li et al. 2023; Rockwell Automation 2023; Woschank et al. 2020).

*F. Additional implications and applications for AI in industry*

With an industrial sector, AI is likely to have numerous applications, as illustrated for aerospace (see Box 1). A major challenge is to ensure that various individual systems are integrated and "talk" to one another. As that challenge is addressed, and AI integration occurs, industrial processes are more likely to be deeply transformed with a multiplication of effects for organizational structures, work tasks, and management.

Additionally, there are many potential applications for AI in manufacturing that are not "shop-floor" specific. For example, automated systems have already been applied in remote customer service, to filter customers to automated instructions or to human representatives (Ameen et al. 2021; Song et al. 2022; Zhao et al. 2022) and in hiring, particularly to automate application rejections (Bongard 2019; Raub 2018; Upadhyay & Khandelwal 2018). Artificial intelligence is also used to evaluate loan applications (Bahrammirzaee 2010; Bhatore 2020) and could potentially be applied to other business decisions (such as evaluating a potential business partner). Such applications could particularly affect small and midsize manufacturers as they seek finance or contracts.

Although it is emerging at scale only recently, there is the possibility that large language models could be used to automate substantial components of the programming process for machine control, automated administration, and other design, managerial and creative applications (Austin et al. 2021; Xu et al. 2022). These are general applications of AI that are likely to have substantial effects in manufacturing, as in other sectors.

III. SOCIETAL IMPLICATIONS OF ARTIFICIAL INTELLIGENCE IN MANUFACTURING

Our discussion of the societal implications of the increasing use of AI in manufacturing draws primarily on the literature that we reviewed (see Appendix 3). Many of these reviewed articles discuss AI as part of an ensemble of technologies, also including, e.g., the "Internet of Things," big data, and additive manufacturing, ushering in a "Fourth Industrial Revolution" or "Industry 4.0." Where possible, we drew on only on those portions of these articles that discussed AI. However, some papers did not parse out the implications of different technologies separately.

The articles reviewed discuss AI's implications for two sets of values and perspectives—those of firms and those of society more generally. Firm-level discussions are largely sanguine, with authors arguing that AI could enhance firms' adaptive capacity, awareness of and responsiveness to customers, and resource efficiency, among other values. One major firm-level concern raised is potential increases in cybersecurity vulnerability due to reliance on AI systems.

Discussions of societal values are more ambiguous and multivalent. Authors suggest that AI would likely contribute to increases in total factor productivity on the societal level. However, they observe that AI in industry could either advance or undercut economic equity, environmental health, national or global security, and political solidarity in a variety of different ways.

In the following, we organize, discuss, and build upon perspectives about implications that we found in the reviewed literature on AI in manufacturing.

*A. Industrial artificial intelligence implications at the firm level*

As noted, articles that address AI's implications for firms' values and operations heavily emphasize potential benefits. This literature suggests that AI could help manufacturers to improve agility and responsiveness on the level of day-to-day operations, e.g., through rapid reconfiguration of designs or manufacturing tools, and on the strategic level, e.g., through reconfiguration of supply chains in response to disruption (Abubakr et al. 2020; Dohale et al. 20222; Park 2017). Some authors explicitly suggest that AI could help to facilitate innovation, e.g., through facilitating product design (Walk et al. 2023; Yang et al. 2020). Similarly, several authors suggest that AI could improve firms' understanding of and ability to respond to customer preferences, through collecting, monitoring, and analyzing use and preference data on customers and by permitting automated and rapid customization of product or service offerings (Abou-Foul et al. 2023; Akinsolu et al. 2023; Caruso 2018; Park 2017).

Reviewed articles addressing the effects of AI on firms' product and service quality take a sanguine view. These papers, mostly from the business literature, suggest that the aforementioned customization opportunities, combined with improved ability to foresee, prevent, and compensate for process and product issues, will allow firms to operate more reliably and to fulfill customer needs more satisfactorily (Abou-Foul et al. 2023; Agrawal et al. 2023; Akinsolu et al. 2023).

Multiple authors argue that AI will permit firms to operate more efficiently, e.g., by improving process reliability and quality and by permitting intelligent planning and real-time adjustment to reduce resource and energy waste (Abou-Foul et al. 2023; Fanti et al. 2022; Gao 2023). However, two papers (Bécue et al. 2021; Iphofen & Hritikos 2021) note that, due to bugs or divergences between AI systems' operating assumptions and actual circumstances, AI systems may sometimes render invalid recommendations or decisions that could result in harm or waste. There is no discussion of whether AIs might make such errors at greater or lesser rates or severity than human beings, though discussants tend to presume that humans still hold the edge over AIs in understanding and responding to unprecedented circumstances.

Only two papers (Akinsolu et al. 2023; Bécue et al. 2021), among the literature we review, explicitly discuss AI's implications for firm security. Both suggest that increased reliance on AI systems in firms will increase their cyberattack surface, i.e., increase the volume of operations that could be disrupted with successful cyberattack. Bécue and colleagues (2021) also suggest that cyber attackers may use AI tools to develop more frequent, adaptive, or powerful cyberattacks. However, they also observe that AI could be used to detect and respond to cyberattacks, potentially resulting in an AI cybersecurity "arms race" between firms and attackers.

## B. Industrial artificial intelligence implications at the societal level

The reviewed articles suggest that AI could have implications for three broad categories of societal values: economic prosperity and equity; environmental health; and community, national, or global security. In aggregate, the articles suggest that, depending upon design and implementation, AI in manufacturing could either support or undercut broad-based economic prosperity for various regions or polities; and contribute to or degrade environmental health. Treatments of AI implications for security, meanwhile, more heavily emphasize novel dangers than novel opportunities (Table 2).

*Table 2. Broad categories of societal values implicated by AI in manufacturing and discussed in reviewed literature*

| Category | Implications |
|---|---|
| Economic Prosperity & Equity | <ul><li>Distribution of wealth & power</li><li>Economic productivity</li><li>Economic resilience</li><li>Job availability & distribution</li><li>Job quality</li></ul> |
| Environmental Sustainability | <ul><li>Resource use</li><li>Waste output</li></ul> |
| Public security & Safety | <ul><li>Cybersecurity</li><li>Dual-use technology</li><li>Legal accountability for harms</li><li>Misinformation</li></ul> |
| Privacy & Autonomy | <ul><li>Personal data collection, aggregation, & analysis</li><li>Manipulative advertising</li></ul> |

*Source*: Authors' elaboration drawing on literature reviewed (see Appendix 2).
*Note*: AI could either advance or undercut societal values in each domain, depending upon design and application.

In the following, we consider these four categories in detail. It should be noted that the literature on AI in manufacturing that considers societal implications represents only a small subset of the broader literature on AI's actual and potential effects on society. As it is likely that manufacturing will also be impacted by these general trends, we also draw (in the subsequent section) on relevant references from this broader literature to develop further insights on societal implications for manufacturing.

Many articles discuss potential implications of AI for economic prosperity and equity, the two of which cannot be neatly separated. Most of these suggest that AI could help to increase both the gross volume of economic activity and total factor productivity, including through reductions in use of labor, material, and energy per unit output; through increases in process speed and reliability; and through increases in manufacturing flexibility and resilience (Park 2017; Szajna & Kostrzewski 2022; Wirtz et al. 2018). AI can permit, and indeed has already permitted, new products and product functionalities (Boninsegni et al. 2022; Caruso 2018; Park 2017). Some authors suggest that AI may help to open new markets, e.g., through dramatic reductions in the costs of some goods or services (Birtchnell & Elliot 2018; Park 2017).

Thus, we find little dispute in the present literature that AI is likely to increase productivity and productive efficiency. However, total factor productivity, as operationalized by economists and firms, never actually includes all the factors of production, i.e., all the scare resources consumed. Caruso (2018) observes that increases to firms' or sectors output-input ratios often are achieved in part through increases in externalized costs, e.g., in consumption of scarce resources, production of waste, or generation of carbon emissions. Thus AI-driven productivity increases may come not only through efficiency increases but through increases in exploitation of human and natural resources.

There is, furthermore, much more to prosperity than the gross volume of economic activity or total factor productivity. Regardless of how much wealth, goods, or services are produced or rendered or at how low a cost in labor or resources, an economy cannot truly be said to be prosperous unless most of its citizens and workers share in that wealth, enjoying safe and meaningful work and leisure, political and economic self-determination, and high standards of living (Castellacci 2022; Mazzucato 2018; Scott 2014). Views on AI's potential implications for jobs and work vary widely. Many articles voice concerns that AI may displace or render precarious many jobs, through automation of routine physical and administrative tasks; and even through encroachment into the domain of design and strategy (Akinsolu et al. 2023; Avis 2018; Birtchnell & Elliott 2018). A few articles argue that, rather than replacing workers, AI systems will work alongside them, increasing their reliability and productivity (Abubakr et al. 2020, Nahavandi 2019). But, on the societal level, the job impacts of job automation and job augmentation may differ only in degree rather than kind. For a given volume of output, increase in labor productivity decreases the amount of labor required, whether through worker augmentation or replacement. Thus, for products with low elasticity of demand, labor required will indeed decrease with increases in labor productivity. Past examples of such changes may be found in fields where human workers have already been heavily augmented by machinery, e.g., manufacturing and agriculture.

There is debate in the literature about whether instances of technological unemployment will be temporary and offset by newly created employment. Certainly, new technologies add as well as eliminate jobs. But job automation and augmentation change the skill demand mix in the economy. Several articles suggest that AI will add computer science, engineering, and maintenance jobs (Abubakr et al. 2020; Dwivedi et al. 2021; Leigh et al. 2020; Xie et al. 2021). But persons displaced by AI systems likely will not possess the skills to win these jobs without substantial retraining. Even if AI maintains or increases labor demand, the types of labor demanded will shift. The most recent wave of job automation has contributed to a job polarization in many industrial economies, with solid middle-class jobs "hollowed out" while demand for high-paying, high-skill, and for low-paying, precarious jobs both grow (Acemoglu & Autor 2010; Autor 2015). Several articles suggest that AI may continue this trend (Avis 2018; Caruso 2018; Park 2017; Wirtz et al. 2018). Some articles explicitly argue that AI is likely to deskill some tasks or jobs (Szajna & Kostrzewski 2022; Xue et al. 2022), while others argue that it will require or facilitate worker upskilling (Avis 2018; Dwivedi et al. 2021; Gao 2023). In a rare empirical examination, Xue and colleagues (2022) find that AI applications in Chinese firms have increased employment of non-college-educated workers and decreased employment of college-educated workers (though these effects were stronger in service than in manufacturing industries).

AI's implications for job quality are similarly multivalent. Lower-skilled jobs might (or might not) be easier, but are also likely to be less satisfying, less well remunerated, and more precarious. Several articles voice concerns about the nature of work with AI. AI monitoring and control might be used to tighten supervision, control, and exploitation of workers, both through direct monitoring and through fragmentation of labor forces. (Caruso 2018; Dwivedi et al. 2021; Fanti et al. 2022). Wirtz and colleagues (2021) argue that workers in "data-rich" environments, tracked and recorded by AI systems, may be exposed to substantial privacy risks. A few authors suggest that persons displaced from "hands-on" work to supervision of automated systems, or from interacting with humans to interacting with robots, may suffer from alienation from their work and from society (Avis 2018; Wirtz et al. 2018). However, some authors also observe that AI could be used to increase job quality and worker autonomy, in particular, by facilitating flexible and worker-led scheduling and workload assignation (Caruso 2018; Nahavandi 2019). Abubakr and colleagues (2020) argue that AI could be used to improve job safety.

The reviewed articles also show no consensus on the likely effects of AI on the distribution of economic opportunities. Caruso (2018) observes that some AI advocates suggest that AI systems may help older workers to remain current and to extend their working lives, which might offer such workers greater opportunity but also might offer justification for the withdrawal or postponement of retirement benefits or social support. Hammer and Karmakar (2021) suggest that AI is likely to intensify demographic inequities in economic opportunity by eliminating mid-skill jobs and increasing the importance of education. Avis (2018) and Park (2017) suggest that AI-driven productivity and flexibility improvements could facilitate "job reshoring" in advanced economies that have suffered job offshoring over the last half-century. Reshoring could benefit workers in wealthy nations but also reduce opportunities for those in poorer nations.

More broadly, the reviewed literature offers diverse but generally negative views on the implications of AI for distribution of economic power. Park (2017) alone suggests that AI-driven coordination may support small-firm competitiveness, and four papers (Birtchnell & Elliott 2018; Boninsegni et al. 2022; Caruso 2018; Dwivedi et al. 2021) suggest that, by reducing the cost of labor, AI may reduce the barrier to entry for firms in certain fields. But, in contrast, several papers voice concerns that AI may increase large firms' economic advantages and contribute to monopolization or oligopolization, e.g., by increasing the marginal value of capital investment advantages and by offering great returns to firms that possess or can acquire large datasets (Caruso 2018; Dwivedi et al. 2021; Fanti et al. 2022; Wirtz et al. 2018). Several papers suggest that AI technologies will contribute to continued wealth polarization, both between citizens in advanced economies and between rich and poor nations, largely by offering opportunities for income or production increases accessible only to the wealthy (Akinsolu et al. 2023; Dwivedi et al. 2021; Iphofen & Hritikos 2021). Caruso (2018) observes that rhetoric around the "Fourth Industrial Revolution" suggests a shift from physical to knowledge capital as the basis of wealth, but that this latter, too, is a scarce and inequitably distributed resource. Avis (2018), meanwhile, suggests that AI technologies may facilitate rent-seeking behavior and "financialization," i.e., pursuit of profits without engaging with production at all. Thus, they may contribute to diversion of economic resources away from productive activity.

In short, reviewed articles agree that AI is likely to increase production but disagree about who is likely to capture the benefits of such increased production. Caruso (2018, p. 390) summarizes the uncertainty well:

> As it has always occurred in the history of the relationship between capital and labour, the possibility that the production process will shift in a direction favourable to labour mainly depends on the capacity for coalition and conflict and on the bargaining power of the latter. These elements develop within the labour relationship also thanks to the support of dynamics (political, cultural, organisational) and actors which are external to the production process, as the history of the workers' movements demonstrates (Bartolini 2000). Therefore, positive outcomes of 'Industry 4.0.' for workers will depend on social conflict and politics.

Concern about AI's implications for environmental sustainability has grown recently, as AI has become not only more widely used but also increasingly computing- and data-intensive. Yet there is no consensus about the environmental sustainability consequences of industrial AI. Indeed, some papers suggest that AI-supported efficiency improvements may decrease use of energy and natural resources (Laskurian-Iturbe et al. 2021; Lei et al. 2023; Yang & Shen 2023). However, steady historical efficiency gains in industry and consumer applications have not resulted in lower aggregate energy consumption, but merely increased output-input ratios. Indeed, in the extreme case, efficiency gains may incentivize increased consumption, a phenomenon known as the Jevons paradox (York & McGee 2016). Meanwhile, several papers observe that AI technologies may drive increases in extraction and use of scarce mineral resources, use of energy for computing, and production of electronic waste (Abubakr 2020; Caruso 2018; Hoosain et al. 2020).

While attention to public security and safety, privacy and autonomy is less evident in our set of reviewed articles, there is some attention to these aspects. Going beyond the more documented risks of cyberattacks on individual firms, Akinsolu and colleagues (2023) voice several concerns about AI's security implications. They observe that an industrial society dependent on computing and AI is more vulnerable to cyberattack and disruption, and furthermore that contemporary information technologies facilitate espionage. Levy (2018) aims to draw out the political consequences of industrial AI, suggesting that AI-driven job losses may drive continued populism in American politics and vulnerability to authoritarian appeals. Boninsegni and colleagues (2022) and Dwivedi and colleagues (2021) observe that AI decision-making renders decision responsibility more ambiguous and raises legal questions about who should be held accountable for AI-inflicted harms, e.g., shop-floor or vehicular injuries. These concerns are equally applicable to more diffuse and more systemic harms, e.g., sale of a product that is defective or that deals harm to consumers, or cost-cutting efforts that leave many people unemployed.

## IV. Societal implications: What's included and what's missing?

Although the current literature on the societal implications of AI in industry covers a large amount of ground, it focuses more deeply in some areas than in others. Of the reviewed articles, 34 discuss economic implications of AI, while 18 address environmental implications and only three address security and public safety. Thirty-six discuss potential benefits of AI in manufacturing, while 20 discuss potential dangers. Twenty-five discuss AI's implications for economic productivity and 21 its implications for jobs and work, but only 10 discuss AI's implications for the distribution of wealth and power in society. To date, there is only limited investigation of the potential implications of AI in manufacturing for public safety and security, for environmental sustainability, and for the structure of the economy and socioeconomic equity.

Moreover, the treatment of certain aspects tends to overlook alternative possibilities. For example, discussions of AI's implications for economic productivity, as well as AI's implications for firms, invariably take a positive view, save for some modest concerns about cybersecurity. We wonder: What are the possibilities that AI could harm the operations of firms or economies, or at least some parts thereof? Could pervasive AI applications that seem beneficial for individual firms generate risks when aggregated and interrelated at an economy- and society-wide level? Previous rounds of automation and offshoring have harmed the manufacturing economies of the United States and European nations even as gross domestic product has grown (driven primarily by services). Prior aspirations toward informed, scientific management and structured coordination have led to economic fragility and human disasters (McNeill 1982; Scott 1998). Financial decision-making based upon sophisticated quantitative models combined with under-regulation of the financial sector to produce the 2008 housing market crash in the United States, and the resultant global recession (Allen & Carletti 2010; Moulton 2014; Immergluck 2015). Some authors argue that the same roots have diverted U.S. firms' investment and attention from enhancing human and physical capital to mergers, acquisitions, selloffs, financial juggling, and executive bonuses over the last half-century (Mazzucato 2018; McCraw & Childs 2018).

While it is impossible to accurately predict the future, it is possible – and essential – to undertake anticipation of potential future pathways and to reflect on, and engage with, contrasting scenarios. In this regard, both optimistic and critical perspectives are needed to fully assess the stakes and possible outcomes of AI in manufacturing. Over the remainder of this section, we use related literature and historical examples to discuss several possible societal effects of manufacturing AI underdiscussed or neglected in the reviewed literature. We do not expect this to be a comprehensive accounting of "missing issues"—indeed, we believe that a comprehensive accounting of issues epistemically requires input from all sectors of society—but we believe we have identified several important gaps.

### A. Appropriation or decreased transferability of knowledge as a threat to innovation and prosperity

Generative AI mimics and recombines elements of preexisting designs. At present, it provides no citations to the designs it mimics. Artists and photography IP-holders whose publicly available, but not public-domain, work has, without license or consent, been used to train image-generating AIs have already launched lawsuits against AI system developers (Appel et al. 2023; Chen 2023). Similar problems could emerge in generative design, with AI systems appropriating the work of designers without citation or license.

The possibility of AI appropriation of the products of human authorship or inventorship introduces several potential disruptions to the current balance of intellectual property law. This is particularly relevant for manufacturing, which is patent intensive and accounts for a majority of all US patents (Fort et al, 2022). The purpose of intellectual property is to encourage innovation and knowledge-sharing by allowing inventors (and other creators) fixed-term monopoly rights in exchange for making public their works and innovations. If AI design goes unchallenged and threatens such monopolies, being able to "remix" or imitate recent inventions without citation or license, firms and individuals may be disincentivized to generate or share inventions (or designs and similarly imitable products). Meanwhile, if AI-generated designs are themselves held to be unpatentable (as is presently the case in the European Union, United Kingdom, and United States; George & Walsh 2022, Villasenor 2022), this may chill investment in or use of AI tools for design and foreclose any advantages they might bring.

If efforts to permit patenting of AI-generated designs progress, this also raises questions about to whom patents should be assigned, given the wide variety of agents and multiple firms likely to be involved in building, training, and implementing AI systems; and the varying degrees of human guidance which AI systems can be given. If the firms that own or operate AI design tools alone receive patents, the human inventors and workers who (willingly or unwillingly, wittingly or unwittingly) support, operate, and produce data and knowledge inputs to such tools may not be appropriately remunerated for their work and may be disincentivized from pursuing invention. Additionally, broad or overly abundant AI patents assigned to large firms (who have the resources required to develop massive training sets) may thwart competition and innovation and disadvantage SMEs. Achieving a fair and effective balance between the different interests and public values at play in the intersection of AI and intellectual property will require substantial attention and broad sensitivity.

In a different but related vein, much economic, and particularly innovative, activity depends upon interorganizational knowledge flows through the media of personnel migration and informal connections between workers in the same industry (Allen 1977; Collins 2010; Crane 1972). If, in the future, a larger proportion of advanced technical work is done by AI, and if more knowledge is embedded within AI systems and less relevant knowledge travels from firm to firm through employee turnover or through informal interchange, innovation or even upkeep of present performance may be impaired. This possibility is not inevitable—the U.S. National Institute of Standards and Technology (2020) is explicitly concerned to promote sharing of knowledge across and about AI systems between organizations—but it is a meaningful risk, nonetheless.

*B. Goal narrowing as a threat to prosperity, resilience, and environmental sustainability*

Since the 1970s, the United States and many other developed nations have seen a reorientation of corporate governance from production of broad-based and sustainable profit to short-term maximization of shareholder value (McCraw & Childs 2018). The narrowing of goals from long-run creation of value to short-run financial gain to shareholders has contributed to a variety of harmful and risky behaviors, including job offshoring, predatory acquisitions, frequent stock buybacks, and investment in financial products rather than in physical and human capital. Among other factors, a narrow focus on near-term profit led to the offering of risky financial products such as those that led to the subprime mortgage crisis of 2008 (Mazzucato 2018).

However, will current modes of specialized AI systems have cognizance of their actions' consequences for things and events other than their programmed corporate targets? If designed to prioritize cost minimization or profit maximization, an AI decision agent could do so without regard to unexpected implications or additional consequences. This could, in certain circumstances, yield harmful or irresponsible behavior. Unless explicitly instructed to do so, an AI system asked to initiate cost-saving or profit-increasing measures might do so in ways that threaten product quality or harm communities, supply-chain contractors, consumers, or the environment. An AI system (again, unless so instructed) would not necessarily be sensitive to moral or geopolitical considerations, reputational and regulatory risks, or societal ramifications and the long-run development of innovation ecosystems. Even if humans were not entirely removed from decision loops, AI recommendations might allow them to abdicate responsibility for potentially harmful decisions. Humans might be more inclined to accept a socially or environmentally harmful decision suggested by an ostensibly objective AI (Gill, 2020).

*C. Optimization, elimination of redundancy, and increased interdependence in operations*

The literature on AI in manufacturing that we reviewed tends to presume that AI-enabled process optimization and control will invariably lead to higher quality and reliability in industry. However, in production, there is an inherent tension between quality and reliability, on one hand, and cost, on the other. Although AI tools might reduce this tension, aggressive, AI-enabled cost-cutting might nonetheless result in quality or reliability sacrifices. History suggests such sacrifices are a very real possibility. For example, customers are often dissatisfied when dealing with "interactive voice response," i.e., the robotic phone call answering systems that many firms now use to cut the costs of customer service (compare Hacikara 2023, Miller et al. 2011).

More generally, Scott (1998) observes that the maintenance and operation of domains of "formal order," such as firm routines and operations, are always dependent upon informal behaviors including management of process exceptions, accommodation to changing external circumstances, workers acting outside of their formal duties, and even direct contravention of formal plans or directives when the situation calls for it. To the extent that AI systems eliminate such informal accommodations or replace the workers that perform them, they might increase the brittleness and fragility of firm operations.

Similarly, the reduction of redundant activity within firms or within the manufacturing sector more broadly, which the U.S. National Institute of Standards and Technology (2022) suggests as one use of AI systems, could also reduce resiliency within firms. What might seem to an AI system as a redundancy could be useful scale-up or knowledge exchange capability. In addition, redundancies help organizations to cover for process interruptions and to respond to external shocks (Perrow 1984; Roberts 1990a,b; Scott 1998). Moreover, the implementation of artificial intelligence tools by firms, unless wholly developed in-house, may increase their dependency upon other organizations (if, for example, a manufacturer lacks the expertise or permissions to maintain or modify the AI tools it uses). Such increased dependency could reduce adopter agility. Dependency could also render AI adopters vulnerable if vendors provide poor service, are crippled by a cyberattack, or go out of business. Automated systems could also endanger reliability or resilience by altering human behavior. Human operators might rely excessively on automated systems, lose situational awareness, or fail to notice or cover for automated system failures—a phenomenon that has been discussed, for example, in monitoring and aircraft automation (Antonovich 2013; Endsley 1996). As discussed above, AI systems could also result in diffusion of responsibility among human operators, again potentially leading to distraction or to an overly narrow accounting of the potential consequences of strategic or operational choices.

Although the literature on AI in manufacturing typically highlights its reliability, highly automated systems are not infallible, although they may fail in ways different from older, human-dominated systems. Unalloyed reliance upon AI design and decision making in high-stakes contexts could produce unanticipated major errors or even disasters (for example, in aerospace engineering contexts); widespread reliance in lower-stakes contexts could produce more subtle, systemic problems. Many such problems emerge from the inevitably partial and incomplete nature of datasets, which can result in problems of algorithmic bias in systems trained upon them. Humans can (sometimes) adjust for data gaps or biases, but AI systems do not necessarily recognize such.

For example, Christian (2020) discusses an experimental hospital AI system built to recommend whether pneumonia patients should be kept overnight, based on personal characteristics and subsequent course of disease for historical patients. In the historical database on which the AI was trained, patients with a history of asthma were always kept overnight. But the AI system recommended that patients with a history of asthma should be sent home, because they typically recovered—thanks to the inpatient care that the AI recommendation would deny them. Direct analogies in manufacturing could entail an AI system recommending less maintenance for a system that consistently performs well thanks to its current maintenance schedule, or relaxed safety protocols around a machine that causes no injuries because of its extensive safety protocols.

The pneumonia patient example is an error obvious to, and correctable by, an attentive, knowledgeable, and empowered human supervisor, but many other and subtler ones are possible. There is a danger that, in high-volume automated decision-making with emergent and dynamic decision rules, or in organizations with time-limited oversight of automated systems, elusive or even serious mistakes emerging from data partiality could easily slip by. AI systems are, indeed, capable of emergent, unforeseen, and sometimes bizarre forms of error. In manufacturing, these could include misguided recommendations, or, if autonomous, decisions, in purchasing, production ordering and scheduling, fault detection and attribution.

Even in the absence of emergent errors, AI systems are necessarily trained on historical data and cannot immediately recognize or adapt to dramatic changes in market or societal situations. It is imaginable that, at the 2020 outset of the COVID-19 pandemic, an AI procurement system would have recommended against purchasing important inputs to firm operations, expecting elevated resource prices to relax soon. A firm that, at AI direction, bet on supply relaxing in mid-2020 would not have fared well. An AI system would not necessarily recognize the dramatic, and likely long-lasting, disruption of the pandemic for international trade and supply chains. In contrast, a human decision-maker might more likely have recognized the unprecedented nature of the economic situation at the onset of the pandemic and improvised a more appropriate operational strategy. In short, neither humans nor AIs are error-free, and both have biases based on their experiences or data, as well as more mysterious details of cognitive structure. However, humans are still more flexible than AI systems, capable of reframing problems, recognizing novel circumstances of relevance, and contextualizing interpretation of data. Moreover, human organizations are well acquainted with human frailties and accustomed to detecting and compensating for them. AIs can err in new, strange, and sometimes more dramatic ways compared to humans. If organizations place too much faith in them, either by uncritically accepting their recommendations or simply by leaving them without human review or oversight, novel sorts of catastrophes may ensure.

*D. Algorithmic social bias as a threat to equity*

The examples above illustrate the ways in which artificial intelligence systems trained on necessarily partial historical data or designed to pursue narrow definitions of performance could, as functions of these biases, impair firm operations or produce general societal harms. However, far more widely discussed in general AI ethics literature is algorithmic social bias. Concerns about algorithmic bias stem from the fact that computerized systems trained on biased historical data sets – often data produced by white, male, English-speaking residents of the Global North – will tend to produce biased outputs that reinforce and exacerbate historical injustices (Benjamin 2019; Danks & London 2017; Hall & Ellis 2023; Kordzadeh & Ghasemaghaei 2022).

Algorithmic bias can be intentional, but typically is not; rather, it can be produced through a combination of the unquestioned or unrealized background assumptions and foci of technology developers, the users in mind when systems are developed, and the historical data with which systems are trained. Algorithmic bias has emerged in many domains. Amazon's (now scrapped) AMZN.O machine learning recruitment algorithm inferred from male predominance in the tech industry that men are better candidates for tech jobs (Dastin 2018; Goodman 2018). Algorithmic tools used to assess the risk of recidivism in criminals or of loan nonrepayment in applicants have, in the United States, yielded longer sentences and lower loan approval rates for Black than for similar white persons (Angwin et al. 2016' Biddle 2022). Algorithmic predictions of where crimes will occur can lead to over-or under-policing of certain communities (Babuta & Oswald 2019; Brantingham 2018). Some facial-analysis tools used to identify persons and to assess characteristics such as age, mood, and gender have been found to be dramatically more accurate for light-skinned men than for dark-skinned women (Buolamwini and Gebru 2018; Hardesty 2018).

These examples arise outside of manufacturing, but it is conceivable that similar (or new) algorithmic biases could emerge in the manufacturing sector. There are at least three ways in which this could occur. First, AI systems are already being applied to hiring in industry, particularly to sift through applications. Such tools can reproduce existing systemic or institutional biases, typically against women or persons of color (Bogen & Rieke 2018; Raghavan et al. 2020). These biases are as concerning in manufacturing as in any other field of hiring. Furthermore, many authors suggest that AI could be used for other contracting purposes, e.g., strategic decision-making, identification of vendors or potential business partners, and analysis of market and demographic preferences. AI tools might reproduce existing racial, gender, physical, locational, and economic biases. It is imaginable, for example, that an AI system trained on biased, historical datasets might demote firms in historically depressed locales or headed by racial minorities or women; or overlook certain market segments (say, persons with unusually large or small bodies).

Second, if computer vision systems are weaker at detecting dark-skinned persons, wheelchair users, or other persons whose bodies may not be included in training datasets, such systems might fail to warn such persons of danger or might offer invalid instructions at higher rates. They also might misapprehend or even misattribute such persons' behavior, leading to inappropriate or misplaced guidance, disciplining, rewards, or time tracking.

Third, AI could be used to identify and prescribe work practices and routines that raise scientific management to the level of hyper-Taylorism. This includes the automated micro-monitoring and management of work and workers at a level of detail unavailable to Fredrick Taylor when he pioneered the measurement of worker efficiency in industry in the early twentieth century. Such detection and prescription could fail to take account of task-relevant differences in persons' bodies, for example, height, weight, limb proportions, fitness, health, age, and physical ability. Thus, they could potentially prescribe work processes that are ineffective for or even injurious to persons with non-modal bodies. Yet, highlighting the importance of responsible technology shaping at early design stages, AI could also potentially be used to customize workflow for the particular bodies and capabilities of different workers, and, thus, to make work processes better fitted to workers and more inclusive.

*E. Loss of privacy or autonomy*

Artificial intelligence in industry also raises concerns of privacy and of manipulation (Manheim & Kaplan 2019). Automated AI technologies have already been applied to microtargeting advertisements and social media feeds, based upon data collected on individual citizens—raising concerns about privacy and autonomy, especially when large amounts of personal data from multiple sources is harvested and inter-linked. There has been much attention on concerns raised by such practices for politics, election fairness, democracy, personal autonomy, and consumer decision-making (Heawood 2018; Ma & Gilbert 2019; Rosenberg et al. 2018). However, politics is not the only domain in which manipulative advertising is possible. Video game companies have developed a wide variety of mechanisms to encourage purchasing of in-game items, including gambling-like "loot boxes" and a multiplayer matchmaking system intended to encourage players to purchase cosmetic items owned by higher-skill players (Alexandra 2017a,b; Marr et al. 2019).

The possibilities of privacy violation and of manipulation are not unique to artificial intelligence. They spread across all digital technologies that collect data about human beings. Nor, perhaps, are privacy problems as obvious in manufacturing as in domains more explicitly concerned with the collection, analysis, and sale of data about humans, e.g., healthcare, advertising, or social media. However, AI in manufacturing could nevertheless threaten privacy and autonomy in two major ways. First, AI could, as discussed above, be applied to surveillance, tracking, and analysis of workers. Data could be used to appropriate workers' skills and task innovations, enforce tighter work discipline, identify labor, equity, or inclusion advocates, or profile workers' health and productivity. In principle, such data could be used for workplace discrimination or sold to other firms; and its collection and analysis could constitute a violation of workers' privacy or organization rights on its own. Second, AI could be used to manipulate workers or nudge them without their knowledge or consent, e.g., to work longer shifts or accept lower pay than they otherwise would (Scheiber 2017; Susser, Roessler, and Nissenbaum 2019). Third, some authors suggest that AI could be used for analysis of customer use and data, and for targeting products to customers (Brunton et al. 2021; Burström et al. 2021; Sahoo & Lo 2022). Microtargeted advertisement is a double-edged sword. On the one hand, with robust data protection and regulation, it could result in more cost-effective marketing. On the other hand, lacking such procedures and their enforcement, large-scale data collection of consumer and user information could further reduce individual privacy and facilitate unethical and nefarious uses (including identity theft).

*F. Dual-use artificial intelligence*

Artificial intelligence technologies are often characterized as "general-purpose" technologies, i.e., as technologies, like steam power, electricity, and computers, applicable to a very wide array of potential uses (Crafts 2021). Thus, they are inherently "dual use," i.e., applicable to both civilian and military ends. AI development for manufacturing applications will likely build human capital, organizational capacities, and technical knowledge that could facilitate development and implementation of military AI (and vice-versa). For example, computer vision tools for tracking workers on factory floors could be reapplied to (or adapted from) weapons targeting tools. Tool manipulation programming could support (or be built from) weapons handling programming.

There is debate about the desirability and ethics of using AI to manage and target offensive weapons. Additionally, employees in some high technology firms have raised concerns about their companies being engaged in military contracts. At the same time, there are many potential defense applications of AI, not only for weapons systems and intelligence but also for enhancing the defense-industrial supply base, managing logistics, and training. The U.S. and many other nations are already investing in military AI (Morgan et al. 2020; Svenmarck et al. 2018), while agencies such as the U.S. Defense Advanced Research Projects Agency (DARPA) have been at the forefront of supporting early research and experimentation which has led to the development of many technologies relevant to information processing, AI, AI-enabled manufacturing, industrial automation, and autonomous systems. The transfer of (taxpayer-developed) R&D and technologies to the civilian industrial sector is generally viewed as desirable. The effective and responsible use of AI-enabled and civilian-developed industrial capabilities is also critical for the defense-industrial base.

The key point is not that either civilian or military AI will necessarily lead to the other, but rather that it will be very difficult to have one without the other. Manufacturers engaged in AI-enabled manufacturing that has military applications will need to consider if and how lines can be drawn which justify or limit such applications. Importantly, employee voices should be considered. Conversely, applications of military-developed AI in civilian manufacturing should be reviewed to ensure that they are relevant and adjusted to civilian requirements and societal considerations (including those related to bias and data sharing).

Defense AI is not the only dual use facilitated by civilian AI development. Improvements in AI technologies may also support improvements in disinformation targeting, cyberattacks, border control, and counter-terrorism among other applications. In addition to AI software, such applications may also require hardware manufacture (including cameras, other sensors, satellite systems). Again, manufacturers will need to face and weigh challenges of risk reduction and security with those of ethics, responsible innovation, and the protection of individual and civil rights in designing such applications.

*G. Further (and unanticipated) implications*

The discussions in the above section attempt to highlight societal implications that are not raised or are underemphasized in the current literature on AI in manufacturing. Even with these additions, this paper makes no claim to have identified all the ways in which industrial AI could potentially affect societal values.[2] Moreover, it is the nature of early-stage technological development that implications will arise that are unanticipated and will not be realized until subsequently, when the technology is widely diffused (Collingridge, 1980). There is a real problem of "lock-in" here in that, as AI becomes more intensively developed and widely diffused, it will become more difficult to make subsequent changes. This stresses the importance of real-time and early assessment of societal implications. In the case of AI in manufacturing, the elucidation of further implications will require consistent attention to the ways in which AI is being implemented and the perspectives of a comprehensive set of stakeholders about how AI is affecting them or may in the future affect them.

---

[2] We invite readers of this pre-print to highlight to us applications and societal implications of AI in manufacturing that we have overlooked or misinterpreted.

## V. Concluding discussion: can public values be secured in an uncertain AI future?

This paper has focused on literature related to AI in manufacturing to identify anticipated applications and societal implications. We have made some reference, albeit incomplete, to the much larger literature on the broader implications of AI across all aspects of society. We see both commonalities and variations between our specific case of AI in manufacturing and the wider applications of AI in society.

In the scholarly literature that was identified through our search, we find a high level of consensus that applications of AI in manufacturing will generally be positive for firms (with cyber security raised as a key risk). We have raised a question of whether there should be greater critical scrutiny of potential drawbacks and issues in the use cases made for AI in manufacturing. On the societal implications, there is a more diverse range of perspectives on AI in manufacturing's implications for economic prosperity and equity, environmental sustainability, public security, privacy, and autonomy. We also noted, except perhaps for employment impacts, a relatively low level of attention in the literature reviewed to the societal implications of AI in manufacturing. However, what is available on societal implications illustrates the uncertainty, multivalence, and plasticity of sociotechnical futures. Technologies, particularly early in their development and deployment, are flexible, amenable to design and implementation to pursue different pathways and values. AI is especially plastic. On the factory floor, it could be used to replace workers or to augment their capabilities. It could facilitate more autonomous work processes for workers or surveillance and micromanagement of worker behavior. On the management level, AI in manufacturing could also replace or enhance many tasks. It could eliminate many management and administrative tasks, while also adding new tasks. It could be used to recommend robust, resilient, and sustainable procurement and sales processes, or to optimize short-run profit and, likely, increase fragility and negative externalities.

On the societal level, AI could replace both skilled and unskilled tasks and jobs, help to educate and advance workers, or both. AI could reduce labor demand, thus increasing unemployment and job precarity, or increase the abundance and features of material goods while maintaining steady employment levels. AI could reduce costs for and facilitate decentralized cooperation between small firms or offer further advantages to large and data-rich firms, intensifying monopoly or oligopoly. And AI could decrease energy use or lead to continued increases in energy use, among many other potentials.

Which of these possibilities are realized – and the tradeoffs among them – will depend to a large extent upon the ambitions and goals of the actors who develop and deploy AI manufacturing systems. Some of these potentialities are more technically feasible than others. But the main determinants of AI's societal effects will be the interests and values represented in the research, business, and government ecosystems that develop and apply AI (Klein & Kleinman 2002; compare Nelson et al. 2021). If a single or small set of stakeholder groups—e.g., big companies or AI engineers—dominates AI development, implementation, and governance, AI systems will advance that group's interests and values to the exclusion of others (Palladino 2023).

Of course, almost every major technological development and economic change has had substantial "unintended consequences" and emergent effects. But unintended does not mean unforeseeable or inevitable (Guston 2014; Nelson et al. 2021; Nelson & Selin 2023). Rather, technological development often brackets, occludes, or neglects the values or interests of persons who are not the core developers and drivers of new technologies.

Efforts to broaden the values considered and prioritized in AI development and implementation will inevitably be piecemeal and imperfect. It would be easy to say, for example, that AI development and implementation should not occur at all without the support of a broad societal consensus, but it is difficult to say how such a consensus might be achieved or discerned (compare Brokowski 2018, on a different emerging technology). Moreover, AI technologies have already been extensively developed and increasingly deployed (see, e.g., Deloitte China 2020).

If (and as) AI proceeds, goals and plans for AI technologies should not be set only by scientists, engineers, and research funders, but through engagement and deliberation of all stakeholder groups likely to be affected, including workers and residents in firms and communities where AI technologies will be developed and deployed, or from which resources will be extracted to build, maintain, and power them; and representatives of communities whose national, economic, or resource security or resiliency may be implicated by AI development and implementation.

Open and responsible governance of AI (in manufacturing as well as for society at large) will depend in part upon the willingness of researchers, engineers, and funders building AI systems to take responsibility for recognizing, respecting, and addressing other actors' needs and values. Governance will also depend upon the opportunity and ability of regulators, workers and labor organizations, non-governmental organizations, and other actors outside the technical core to shape AI development and deployment in support of societal needs and values. Methods by which technologists may examine the potential societal implications of their technologies include societally oriented scenario planning, public engagement, collaboration with civil society, government, and social or humanistic scholars, and technology assessment, among others. Methods by which non-technologists may shape technological development include public policy tools such as responsible research, development, and implementation funding, public procurement, regulation, and standard-setting; as well as civil society tools such as convening of dialogues, publication of reports and position papers, activist investing, voting, and other forms of civic engagement.

It is impossible for any small set of engineers or analysts, no matter how well-informed, to foresee or plan for all the ways in which AI manufacturing might affect societal values. Thus, both AI implementation and governance should be kept iterative, flexible, and attentive both to all implicated stakeholders and to changing circumstances. There are surely costs, in terms of time and resources, to facilitate flexibility and inclusivity. Such costs should be viewed as public and private investments over the long-run to enable AI applications in manufacturing that are not only economically viable, but which also promote social cohesiveness and inclusion and environmental sustainability.

ACKNOWLEDGMENTS

This material is based upon work supported by the Georgia Artificial Intelligence Manufacturing (GA-AIM) Corridor at Georgia Institute of Technology. GA-AIM is in receipt of support from the Economic Development Administration (EDA), U.S. Department of Commerce. Any opinions, findings, and conclusions or recommendations expressed in this material are those of the authors and do not necessarily reflect the views of GA-AIM or EDA.

APPENDIX 1
SEARCH AND REVIEW PROCEDURES

To examine possible applications of artificial intelligence in manufacturing, we conducted a systematic review of engineering and management literature offering typologies of possible AI applications in manufacturing. Between April 4th and 15th, 2023, we searched Google, Google Scholar, and Scopus for reports and papers on grey literature in manufacturing. Search terms included "artificial intelligence," "machine learning," "expert system," "large language model," "neural network," "manufacturing" and lemmatized variants, "application," "review," "consensus report," and "white paper." We examined all (665) Scopus results in the engineering, computer science, decision-making, business, sociology, economic, or arts subject areas, the first 40 Google Scholar results sorted by relevance, and the first 100 Google results sorted by relevance, stopping, in the latter two cases, at these points due to a preponderance of irrelevant results. This procedure was intended to capture potential applications of AI presented both in mainstream scholarly and grey literature and in conferences, which are particularly important mechanisms of communication in engineering fields. Manual review of result titles and abstracts yielded 85 scholarly and 22 grey literature papers discussing AI applications in manufacturing. Manual review of these documents' bodies yielded seven grey-literature and 33 academic papers offering explicit typologies of AI applications in manufacturing, for a total of 40 focal papers (listed in Appendix 2).

We iteratively developed our application typology through several rounds of review and coding. First, we developed a draft general typology via synthesis of several of the most comprehensive typologies offered in the literature, from Arinez and colleagues (2020), National Institute of Standards and Technology (2022), Rauch and colleagues (2021), World Manufacturing Foundation (2020). We then coded the body text of all 40 documents into this draft typology using the qualitative data analysis program NVivo, adding categories as necessary and noting any mismatches between applications and the initial draft typology. Review for combinations of similar categories and application placement yielded our final typology.

To investigate possible societal implications of artificial intelligence in manufacturing, we conducted a systematic literature review of academic literature through Scopus. To maximize the relevance and reliability of the literature reviewed, we limited our review to articles published since 2014 in scholarly journals indexed in the research database Scopus and possessing Scopus CiteScores of 3.0 or higher. Preliminary searchers revealed that literature specifically focused on societal implications of AI in manufacturing is scant, so we more broadly reviewed literature on societal implications of AI that included some mention of manufacturing. On May 13th, 2023, we searched Scopus for articles containing terms related to societal implications of AI in their titles, abstracts, or keywords, and including references to artificial intelligence in their titles or abstracts. We limited results to the environmental and social sciences to focus on papers focusing on societal implications of artificial intelligence.

Our search yielded 154 papers. We manually reviewed the titles and abstracts for relevance, downloading 54 papers for review of full text. Review of full text identified 44 papers that substantially discussed possible societal effects or implications of AI in manufacturing (see Appendix 3). We used the qualitative data analysis program NVivo to inductively code societal implications or effects of AI posited in these 44 papers. We organized implications by the societal values which they affected.

# APPENDIX 2
## AI APPLICATIONS IN MANUFACTURING

The following table lists and categorizes sources (Total N = 40) discussing artificial intelligence applications in manufacturing.

| Application | N | Sources |
|---|---|---|
| ***Supply network*** | ***14*** | |
| *Planning and scheduling* | *9* | |
| Customer targeting | 1 | Burström et al. 2021 |
| Inventory planning | 1 | Bertolini et al. 2021 |
| Market forecasting | 5 | Bertolini et al. 2021; Deloitte China 2020; Nolan 2021; Rockwell Automation 2023; World Manufacturing Foundation 2020 |
| Multi-site or inter-firm production coordination | 2 | Kamble 2022; National Institute of Standards and Technology 2022 |
| Production capacity identification and location | 1 | National Institute of Standards and Technology 2022 |
| *Tracking and monitoring* | *5* | |
| Industry-wide data aggregation and sharing | 1 | National Institute of Standards and Technology 2022 |
| Multi-site inventory, request, and delivery tracking | 4 | Burström et al. 2021; Deloitte China 2020; Kamble et al. 2022; Woschank et al. 2020 |
| *Control and dispatching* | *4* | |
| Ordering and distribution coordination | 4 | Bertolini et al. 2021; Li et al. 2023; Rockwell Automation 2023; Woschank et al. 2020 |
| ***Manufacturing facility*** | ***38*** | |
| *Design and development* | *29* | |
| Process parameter prediction and selection | 25 | Arinez et al. 2020; Bertolini et al. 2021; Gentner et al. 2022; Goh et al. 2021; Huang et al. 2021; Kamble et al. 2022; Khosravani & Nasiri 2020; S. Kumar et al. 2022; S.P.L. Kumar 2019; Li et al. 2023; C. Liu et al. 2022; Meng et al. 2020; Nguyen et al. 2022; Peres et al. 2020; Qin et al. 2022; Rauch et al. 2021; Sahoo & Lo 2022; Sarkon et al. 2022; Sira 2022; Tao et al. 2018; Waltersmann et al. 2021; C. Wang et al. 2020; World Manufacturing Foundation 2020; Woschank et al. 2020; Xames et al. 2022 |
| Search among preexisting process designs | 1 | National Institute of Standards and Technology 2022 |
| Tool selection | 1 | S.P.L. Kumar 2019 |
| *Planning and scheduling* | *21* | |
| Inspection planning | 1 | Brunton et al. 2021 |
| Job scheduling and coordination | 11 | Bertolini et al. 2021; Dogan & Birant 2011; Fahle et al. 2020; Kamble et al. 2022; Li et al. 2023; Rauch et al. 2021; Tao et al. 2018; J. Wang et al. 2022; World Manufacturing Foundation 2020; Woschank et al. 2020; Z. Zhang et al. 2022 |
| Maintenance planning and coordination | 17 | Dogan & Birant 2021; Fahle et al. 2020; Gentner et al. 2022; HPE Greenlake 2022; Huang et al. 2021; Kamble et al. 2022; Li et al. 2023; Mumali 2022; Nguyen et al. 2022; Peres et al. 2020; Rauch et al. 2021; Sahoo & Lo 2022; Sira 2022; Tao et al. 2018; J. Wang et al. 2022; World Manufacturing Foundation 2020; Z. Zhang et al. 2022 |
| Plan assessment | 1 | Huang et al. 2021 |
| Resource allocation | 6 | Arinez et al. 2020; Dogan & Birant 2021; Huang et al. 2021; Kamble et al. 2022; Sahoo & Lo 2022; Sira 2022 |
| *Tracking and monitoring* | *35* | |
| Cybersecurity | 4 | Goh et al. 2021; Nguyen et al. 2022; World Manufacturing Foundation 2020; Xames 2022 |
| Energy and resource use tracking | 6 | Deloitte China 2020; Li et al. 2023; Tao et al. 2018; Waltersmann et al. 2021; World Manufacturing Foundation 2020; Woschank et al. 2020 |
| Industrial espionage | 1 | C. Wang et al. 2020 |
| Maintenance tracking and documentation | 4 | Bertolini et al. 2021; Deloitte China 2020; C. Liu et al. 2022; Rockwell Automation 2023 |
| Output and quality monitoring and attribution | 28 | Arinez et al. 2020; Bertolini et al. 2021; Chinchanikar & Shaikh 2022; Deloitte China 2020; Dogan & Birant 2021; Fahle et al. 2020; Gentner et al. 2022; Goh et al. 2021; HPE Greenlake 2022; Huang et al. 2021; Khosravani & Nasiri 2020; S. Kumar et al. 2022; Li et al. 2023; Meng et al. 2020; Mumali 2022; Nolan 2021; Peres et al. 2020; Qin et al. 2022; Rauch et al. 2021; Rockwell Automation 2023; Sahoo & Lo 2022; Sarkon et al. 2022; Tao et al. 2018; C. Wang et al. 2020; J. Wang et al. 2022; World Manufacturing Foundation 2020; Xames et al. 2022; Z. Zhang et al. 2022 |
| Process fault detection and attribution | 19 | Bertolini et al. 2021; Brunton et al. 2021; Dogan & Birant 2021; Gentner et al. 2022; Huang et al. 2021; Kamble et al. 2022; Khosravani & Nasiri 2020; S. Kumar et al. 2022; Meng et al. 2020; Nguyen et al. 2022; Rauch et al. 2021; Sira 2022; Tao et al. 2018; Waltersmann et al. 2021; C. Wang et al. 2020; J. Wang et al. 2022; World Manufacturing Foundation 2020; Xames et al. 2022; Z. Zhang et al. 2022 |
| Worker behavior tracking | 3 | Brunton et al. 2021; Sira 2022; World Manufacturing Foundation 2020 |
| *Control and dispatching* | *19* | |
| Flexible process (re)configuration | 2 | Li et al. 2023; National Institute of Standards and Technology 2022; |
| Job dispatching | 2 | Arinez et al. 2020; Rockwell Automation 2023 |
| Real-time process adjustment | 9 | Arinez et al. 2020; Bertolini et al. 2021; Chinchanikar & Shaikh 2022; Deloitte Chine 2020; Huang et al. 2021; Kamble et al. 2022; S. Kumar et al. 2022; Rauch et al. 2021; Waltersmann et al. 2021 |
| Resource dispatching | 8 | Arinez et al. 2020; Gentner et al. 2022; Goh et al. 2021; Kamble et al. 2022; Li et al. 2023; Rockwell Automation 2023; Tao et al. 2018; World Manufacturing Foundation 2020 |
| Worker instruction (e.g., through augmented reality) | 2 | Fahle et al. 2020; Nolan 2021 |
| ***Manufacturing machine or tool*** | ***25*** | |
| *Design and development* | *2* | |
| Machine design | 2 | Kamble et al. 2022; S.P.L. Kumar 2019 |
| *Planning and scheduling* | *15* | |

| Application | N | Sources |
|---|---|---|
| Machine motion planning | 3 | Huang et al. 2021; Li et al. 2023; National Institute of Standards and Technology 2022 |
| Predictive error compensation | 4 | Meng et al. 2020; Qin et al. 2022; Sarkon et al. 2022; Xames et al. 2022 |
| Predictive maintenance | 9 | Arinez et al. 2020; Huang et al. 2021; Mumali 2022; Rauch et al. 2021; Waltersmann et al. 2021 J. Wang et al. 2022; World Manufacturing Foundation 2020; Woschank et al. 2020; Z. Zhang et al. 2022 |
| *Tracking and monitoring* | *5* | |
| Machine status monitoring | 5 | Arinez et al. 2020; Bertolini et al. 2021; Huang et al. 2021; Rauch et al. 2021; Rockwell Automation 2023 |
| *Control and dispatching* | *13* | |
| Automated maintenance | 1 | Li et al. 2023 |
| Machine control | 12 | Arinez et al. 2020; Brunton et al. 2021; Deloitte China 2020; Fahle et al. 2020; Huang et al. 2021; Kamble et al. 2022; Li et al. 2023; Z. Liu et al. 2022; National Institute of Standards and Technology 2022; Peres et al. 2020; Rauch et al. 2021; World Manufacturing Foundation 2020 |
| **Product or service** | **26** | |
| *Design and development* | *24* | |
| AI-enabled product features | 1 | Deloitte China 2020 |
| Design property prediction and assessment | 13 | Brunton et al. 2021; Chinchanikar & Shaikh 2022; Dogan & Birant 2021; Fahle et al. 2020; Goh et al. 2021; Huang et al. 2021; Khosravani & Nasiri 2020; Meng et al. 2020; Qin et al. 2022; Sarkon et al. 2022; C. Wang et al. 2020; J. Wang et al. 2022; Xames et al. 2022 |
| Design suggestion | 8 | Brunton et al. 2021; Chinchanikar & Shaikh 2022; Goh et al. 2021; Nolan 2021; Rauch et al. 2021; Sarkon et al. 2022; C. Wang et al. 2020; World Manufacturing Foundation 2020 |
| Generative design | 5 | Chinchanikar & Shaikh 2022; Deloitte China 2020; Rauch et al. 2021; Sahoo & Lo 2022; World Manufacturing Foundation 2020 |
| Product customization | 4 | Goh et al. 2021; C. Liu et al. 2022; World Manufacturing Foundation 2020; Xames et al. 2022; Brunton et al. 2021; Nolan 2021 |
| Search within prior design or experiment records | 2 | Brunton et al. 2021; Nolan 2021 |
| Test planning | 1 | Brunton et al. 2021 |
| *Planning and scheduling* | *4* | |
| Contract assessment | 1 | Burström et al. 2021 |
| Forecasting customer product usage | | Burström et al. 2021 |
| Predictive maintenance | 4 | Brunton et al. 2021; Burström et al. 2021; Kamble et al. 2022; Nguyen et al. 2022 |
| *Tracking and monitoring* | *7* | |
| Customer or user feedback processing and analysis | 4 | Goh et al. 2021; Nguyen et al. 2022; Sahoo & Lo 2022; Tao et al. 2018 |
| Product service data collection and analysis | 4 | Brunton et al. 2021; Burström et al. 2021; Deloitte China 2020; Sahoo & Lo 2022 |
| Product status tracking | 4 | Brunton et al. 2021; Burström et al. 2021; Kamble et al. 2022; Nguyen et al. 2022 |
| *Control and dispatching* | *1* | |
| Product upkeep and handling automation | 1 | Brunton et al. 2021 |

# APPENDIX 3
## SOCIETAL IMPLICATIONS OF AI IN MANUFACTURING

The following table lists and categorizes sources (Total N = 44) identified through the literature search related to potential societal implications of AI in industry.

| Value | N | Sources |
|---|---|---|
| **Firm level** | *20* | |
| *Adaptive capacity* | *9* | |
|   Facilitate innovation (B) | 2 | Walk et al. 2023; Yang et al. 2020 |
|   Increase flexibility & agility (B) | 7 | Abou-Foul et al. 2023; Abubakr et al. 2020; Caruso 2018; Dey et al. 2023; Dohale et al. 2022; Olsen & Tomlin 2020; Park 2017 |
| *Market awareness and responsiveness* | *5* | |
|   Customize products and services (B) | 3 | Bourke 2019; Caruso 2018; Park 2017 |
|   Target customers & understand demand (B) | 5 | Abou-Foul et al. 2023; Akinsolu et al. 2023; Bourke 2019; Caruso 2018 |
| *Product & service quality* | *7* | |
|   Increase process reliability & quality (B) | 7 | Abou-Foul et al. 2023; Abubakr et al. 2020; Agrawal et al. 2023; Akinsolu et al. 2023; Caruso 2018; Dey et al. 2023; Wirtz et al. 2018 |
| *Resource use* | *12* | |
|   Waste, loss, or damage from invalid recommendations, decisions, or actions (H) | 2 | Bécue et al. 2021; Iphofen & Hritikos 2021 |
|   Improve resource & energy efficiency (B) | 10 | Abou-Foul et al. 2023; Abubakr et al. 2020; Caruso 2018; Fanti et al. 2022; Gao 2023; Laskurian-Iturbe et al. 2021; J. Liu et al. 2022; Lv et al. 2022; Park 2017; Walk et al. 2023 |
| *Security* | *2* | |
|   Enhance cyberattack capabilities (H) | 1 | Bécue et al. 2021 |
|   Expand cyberattack surface (H) | 2 | Akinsolu et al. 2023; Bécue et al. 2021 |
|   Enhance cybersecurity capabilities (B) | 1 | Bécue et al. 2021 |
| **Societal level** | *43* | |
| *Economic prosperity & equity* | *34* | |
| *Jobs and work* | *21* | |
| *Job availability and distribution* | *19* | |
|   Displace or render precarious jobs (H) | 18 | Abubakr et al. 2020; Akinsolu et al. 2023; Avis 2018; Bhattacharyya & Nair 2019; Birtchnell & Elliott; Caruso 2018; Dwivedi et al. 2021; Fanti et al. 2022; Foster-McGregor et al. 2021; Gao 2023; Hammer & Karmakar 2021; Iphofen & Hritikos 2021; Leigh et al. 2020; Levy 2018; Majumdar et al. 2018; Park 2017; Wirtz et al. 2018; Xie et al. 2021 |
|   Intensify demographic inequities (H) | 1 | Hammer & Karmakar 2021 |
|   Job polarization (H) | 4 | Avis 2018; Caruso 2018; Park 2017; Wirtz et al. 2018 |
|   Add high-skill jobs (B) | 4 | Abubakr et al. 2020; Dwivedi et al. 2021; Leigh et al. 2020; Xie et al. 2021 |
|   Add low-skill jobs (B) | 1 | Xue et al. 2022 |
|   Reshore production in advanced economies (A) | 2 | Avis 2018; Park 2017 |
|   Advantage knowledge workers (A) | 1 | Caruso 2018 |
| *Job quality* | *10* | |
|   Expose workers to privacy risks (H) | 1 | Wirtz et al. 2018 |
|   Intensify control and exploitation of laborers (H) | 3 | Caruso 2018; Dwivedi et al. 2021; Fanti et al. 2022 |
|   Alienate laborers from work (H) | 2 | Avis 2018; Wirtz 2018 |
|   Improve safety (B) | 1 | Abubakr et al. 2020 |
|   Increase worker autonomy (B) | 2 | Caruso 2018; Nahavandi 2019 |
|   Deskill labor (A) | 3 | Avis 2018; Szajna & Kostrzewski 2022; Xue et al. 2022 |
|   Upskill labor (A) | 4 | Avis 2018; Caruso 2018; Dwivedi et al. 2021; Gao 2023 |
| *Productivity* | *25* | |
|   Improve manufacturing flexibility and resilience (B) | 1 | Dohale et al. 2022 |
|   Increase total factor productivity (B) | 24 | Abou-Foul et al. 2023; Abubakr et al. 2020; Agrawal et al. 2023; Akinsolu et al. 2023; Avis 2018; Birtchnell & Elliott 2018; Boninsegni et al. 2022; Botha 2018; Bourke 2019; Caruso 2018; Clauser et al. 2022; Dwivedi et al. 2021; Fanti et al. 2022; Gao 2023; Hoosain et al. 2020; Li et al. 2023; Nahaavndi 2019; Olsen & Tomlin 2020; Park 2017; Szajna & Kostrzewski 2022; Wirtz et al. 2018; Yang & Shen 2023; X. Zhang et al. 2022a; X. Zhang et al. 2022b |
|   Enable new products & product functionalities (A) | 3 | Boninsegni et al. 2022; Caruso 2018; Park 2017 |
|   Open new markets (A) | 2 | |
| *Wealth & power distribution* | *10* | |
|   Diversion of investment away from production (H) | 1 | Avis 2018 |
|   Increased monopolization or oligopolization (H) | 5 | Avis 2018; Caruso 2018; Dwivedi et al. 2021; Fanti et al. 2022; Wirtz et al. 2018 |
|   Wealth polarization (H) | 7 | Akinsolu et al. 2023; Avis 2018; Caruso 2018; Dwivedi et al. 2021; Iphofen & Hritikos 2021; Park 2017; Wirtz et al. 2018 |
|   Support small-firm competitiveness (B) | 1 | Park 2017 |
|   Lower barrier to entry in certain fields (A) | 4 | Birtchnell & Elliott 2018; Boninsegni et al. 2022; Caruso 2018; Dwivedi et al. 2021; |
|   Reinforce interests of capital (A) | 1 | Avis 2018 |
|   Shift bases of firm competition (A) | 1 | Park 2017 |
|   Shift toward capital intensity (A) | 1 | Avis 2018 |
| *Environmental health* | *18* | |
|   Generically degrade environment (H) | 1 | Avis 2018 |
| *Resource use* | *17* | |
|   Increase energy use (H) | 2 | Caruso 2018; Hoosain et al. 2020 |

| Value | N | Sources |
|---|---|---|
| Increase use of electronics component resources (H) | 2 | Abubakr et al. 2020; Caruso 2018 |
| Facilitate material reuse (B) | 5 | Dey et al. 2023; Hoosain et al. 2020; Lei et al. 2023; Ronaghi 2022; Talla & McIlwaine 2022 |
| Reduce energy & materials use (B) | 13 | Laskurian-Iturbe et al. 2021; Lei et al. 2023; Li et al. 2023; J. Liu et al. 2022; Lv et al. 2022; Park 2017; Ronaghi 2022; Schöggl et al. 2023; Talla & McIlwaine 2022; Walk et al. 2023; Yang & Shen 2023; X. Zhang et al. 2022a; X. Zhang et al. 2022b |
| *Waste output* | *1* | |
| Increase E-waste production (H) | 1 | Caruso 2018 |
| **Security & public safety** | **3** | |
| *Legal accountability* | *2* | |
| Render responsibility and accountability ambiguous (H) | 2 | Boninsegni et al. 2022; Dwivedi et al. 2021 |
| *Conventional warfare* | *1* | |
| Improve weapons systems (A) | 1 | Akinsolu et al. 2023 |
| *Cybersecurity* | *1* | |
| Expand cyberattack surfaces (H) | 1 | Akinsolu et al. 2023 |
| More frequent or powerful cyberattacks (H) | 1 | Akinsolu et al. 2023 |
| *Propaganda* | *1* | |
| Target disinformation (H) | 1 | Akinsolu et al. 2023 |
| **Political solidarity and health** | **1** | |
| Foster populism (A) | 1 | Levy 2018 |

**Key**

| Symbol | Meaning |
|---|---|
| [None] | Heading |
| B | Opportunity; potentially beneficial effect |
| H | Danger; harmful or concerning effect |
| A | Potential effect cannot be easily summarized as beneficial or harmful |